# Electrically induced ferromagnetism in an irradiated complex oxide

Nareg Ghazikhanian[1,2*], Pavel Salev[3], Dayne Sasaki[4,5], Yayoi Takamura[4], and Ivan K. Schuller[1,2]

1. Department of Physics, University of California, San Diego; La Jolla, 92093, USA.
2. Materials Science and Engineering Program, University of California, San Diego; La Jolla, 92093, USA.
3. Department of Physics and Astronomy, University of Denver; Denver, 80210, USA.
4. Department of Materials Science & Engineering, University of California, Davis; Davis, 95616, USA.
5. Lawrence Berkeley National Laboratory; Berkeley, 94720, USA.

*Corresponding author, email: nghazikh@ucsd.edu

**Abstract**

In metal-insulator transition materials, a small perturbation can shift the delicate balance between competing or coexisting electronic phases, leading to dramatic changes of the material's properties. Using $La_{0.7}Sr_{0.3}MnO_3$, a prototypical metal-insulator transition manganite, we show that local low-dose focused ion beam irradiation increases resistivity by several orders of magnitude, converting the ferromagnetic-metal ground state into a paramagnetic-insulator. Surprisingly, we found that applying electric stimuli to the irradiated material induces a non-thermal insulator-to-metal transition, which results in low-power, repeatable volatile resistive switching. Magnetotransport measurements revealed that this voltage-induced metallic phase in the irradiated material is ferromagnetic, exhibiting clear anisotropic magnetoresistance. This work, thus, reports the discovery of a unique material in which the electrical triggering of the electronic phase transition results in the onset of magnetism, in stark contrast to the magnetic order suppression commonly observed in other metal-insulator transition switching materials. We demonstrate that local focused ion beam irradiation provides new and exciting opportunities to engineer electronic and magnetic functionalities that can find practical applications ranging from spintronics to neuromorphic hardware.

**Introduction**

In many materials that exhibit a metal-insulator transition, the electronic phase transition is accompanied by magnetic ordering.[1-3] Consequently, the electrical triggering of the electronic phase transition can enable the control of magnetism.[4-9] Typically, this electrical triggering tends to suppress local magnetic ordering, as is observed with the formation of a transverse paramagnetic barrier in lanthanum strontium manganite, (La,Sr)MnO$_3$ (LSMO), and the formation of paramagnetic metallic filaments in vanadium sesquioxide, $V_2O_3$, and neodymium nickelate, $NdNiO_3$.[10-12] To date, a system in which electrical switching leads to local magnetic ordering has not yet been found. Here, we report the discovery of the onset of ferromagnetism during the electrically induced insulator-to-metal (IMT) switching in irradiated LSMO. As-grown LSMO exhibits metallic-ferromagnetic behavior at low temperatures which transitions to a high temperature paramagnetic-insulating phase. Local, focused, $Ga^+$ ion beam irradiation suppresses the metal-to-insulator transition (MIT) and the low-temperature ferromagnetism, rendering the material insulating throughout the entire temperature range. Interestingly, we found that applying a voltage to these $Ga^+$ ion-irradiated films induces a non-thermal *IMT*, as opposed to the *MIT* electrical switching in as-synthesized LSMO. Unexpectedly, this electrical IMT switching also coincides with the onset of ferromagnetism. This shows that applying electric stimuli induces magnetic ordering in the ion beam modified material. Our findings demonstrate that selective ion irradiation presents an efficient tool for controlling the switching type, switching mechanism, and magnetic states in phase-change materials. Developing new functionalities and reducing energy consumption associated with the electrical metal-insulator transition triggering is an important challenge for electronics applications.[13-16] We show that focused ion beam irradiation may provide a highly controllable tool for engineering new, scalable, and energy-efficient electronic and spintronic devices.

**Results**

For electrical and magnetotransport measurements, 50×100 $\mu m^2$ two-terminal devices were patterned on a 20-nm-thick epitaxial LSMO film (Fig. 1A). A 30 keV focused $Ga^+$ ion beam with a 6.2·$10^{13}$ Ga ions/$cm^2$ dose was used to create a 1-µm-wide irradiated strip stretching across the full device width. According to Transport of Ions in Matter[17] simulations, $Ga^+$ ion irradiation induces atomic displacements throughout the entire film thickness where each incident ion creates ~400 vacancies (Fig. 1B).

LSMO with the La/Sr ratio of 0.7/0.3 exhibits a phase transition from the ground ferromagnetic-metal state to a high-temperature paramagnetic-insulator.[18] The resistance-temperature measurements (Fig. 1C, blue line) show a transition at $T_c$ ~ 340 K in the as-synthesized material before local $Ga^+$ irradiation. Overall, the samples showed properties expected from LSMO, confirming the stoichiometric material synthesis.

After irradiation with a focused $Ga^+$ beam, a stark contrast is observed in the electrical transport properties compared to the pristine LSMO. The irradiated device displays insulating behavior throughout the entire temperature range (Fig. 1C, red line). The log-linear dependance of the conductivity (s) shows typical Arrhenius thermally activated transport with an activation energy of ~110 meV (inset in Fig. 1C). We note that although the longitudinal device dimension (i.e. along the current flow) is much larger than that of the locally irradiated area (~99 µm vs. ~1 µm, respectively), the device resistance appears to be dominated by the irradiated region. This indicates a large local resistivity increase of LSMO after being exposed to a small $Ga^+$ ion dose. A minor resistance anomaly is present at temperatures above 340 K, where the pristine LSMO areas surrounding the local irradiated strip undergo the MIT and their resistance approaches the resistance of the irradiated material. The large resistivity increase in the LSMO device after irradiation is consistent with previous studies of ion-irradiated LSMO.[19,20] Similar changes in thermally-driven equilibrium transport behavior have also been reported in strained LSMO films[21,22], films with varying doping concentrations[18], ultrathin films[23], electrolyte-gated devices[24], and oxygen deficient films[25], suggesting that there are multiple mechanisms which convert the metallic LSMO phase into an insulating one. Possible alternative approaches, e.g. modifying film growth conditions or using global irradiation with varying ionic species, could also achieve similar resistive switching and voltage-induced magnetism effects as we discuss further in this work.

Typically, stoichiometric LSMO shows volatile MIT resistive switching from low-to-high resistance, in contrast to IMT materials such as $VO_2$, $V_2O_3$, $V_3O_5$, (RE)$NiO_3$, etc. that exhibit high-to-low resistance switching.[7,26] The typical I-V curves are highly nonlinear showing a pronounced N-type negative differential resistance (Fig 2A), i.e. a part of the I-V curve where $dV/dI < 0$ above a threshold V, characteristic of metal-to-insulator resistive switching. Previous works attributed this switching mainly to electrothermal effects, where the strong nonlinear dependence of LSMO resistivity on temperature results in Joule self-heating under high currents or voltages.[27,28] Because this switching is primarily mediated by Joule heating, the switching is volatile, i.e. the resistance automatically resets to its initial low value upon removal of the electrical bias. It has been also shown that the MIT switching in LSMO occurs by the formation of a local paramagnetic insulating barrier that splits the ferromagnetic metallic matrix.[10] Thus, the MIT switching in stoichiometric LSMO provides electrical control of the magnetic phase by locally suppressing the magnetic order. Contrary to this, switching in the irradiated LSMO induces magnetic ordering by applying voltage as will be shown below.

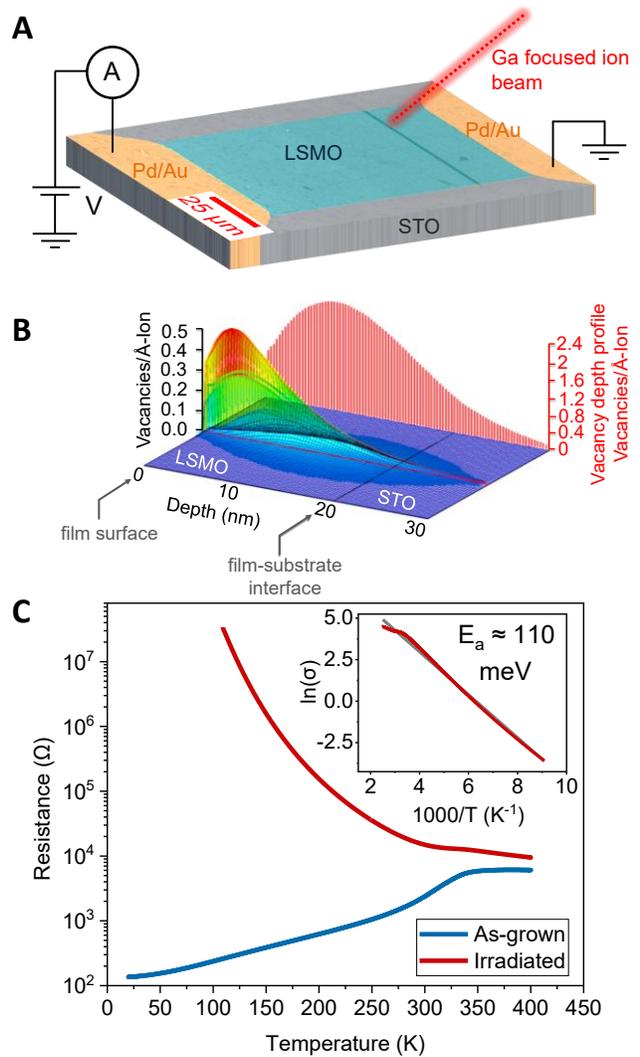

**Figure 1: Irradiation of LSMO (A)** False color scanning electron microscope image of a 50×100 μm² LSMO device. The Ga⁺ irradiated region appears darker in the image (dark blue 1-μm-length strip stretching across the full device length in the false color image). **(B)** Simulated Ga⁺ irradiation modification profile. The red axis displays the total vacancy distribution profile. The surface plot indicates 2D vacancy distribution within 3x3 Å² cells. **(C)** Resistance vs. temperature measurements in the LSMO device before (blue line) and after (red line) focused ion beam irradiation. Irradiation supresses the MIT, redering the device insulating in the entire temperature range. The inset shows a typical Arrhenius plot of the irradiated device conductivity vs. inverse temperature, with an activation energy of ~110 meV.

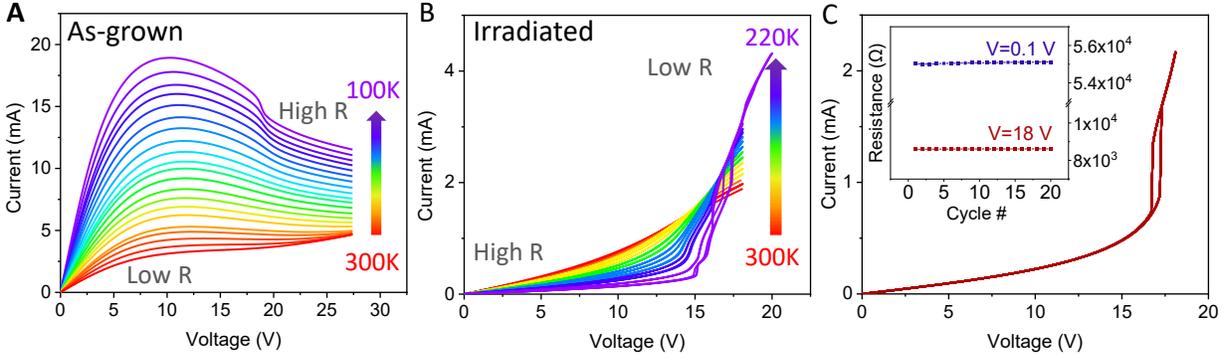

**Figure 2: Change in resistive switching after irradiation (A)** Temperature dependent I-V characterestics of the pristine device shows the expected low-to-high MIT resistance switching of a stoichiometric LSMO. **(B)** Temperature dependent I-V characteristics of the irradiated device showing an unusual high-to-low resistance switching. **(C)** Overlaid 20 I-V switching cycles recorded in the irradiated device at 250 K. The high-to-low switching is highly reproducible. The inset shows the resistance of the device at 0.1 V and 18 V for each switching cycle, exhibiting less than 0.1% cycle-to-cycle variability.

Interestingly, the $Ga^+$ irradiated LSMO devices unexpectedly show volatile *high-to-low* resistance switching, opposite to the *low-to-high* switching in the pristine material discussed above. Fig. 2B shows that the current flowing through the irradiated device initially remains low as the applied voltage is ramped up (i.e. $Ga^+$ ion beam irradiation locally converted LSMO into an insulating state). At a threshold voltage of ~13-15 V depending on temperature, the current rapidly increases, indicating a large decrease of the device resistance, i.e. high-to-low resistance switching. The resistive switching is volatile: i.e. when the applied voltage is removed, the device switches back into the initial high-resistance state. The switching is more pronounced at lower temperatures and even shows hysteretic behavior below ~260 K. At the same temperature (220K) and voltage (16 V) the switching ratio of the irradiated device (1100%) is much larger than the pristine one (277%). Moreover, the switching in the irradiated device is highly repeatable. Fig 2C shows 20 overlaid consecutive switching cycles recorded at 250 K. All the I-V curves are virtually indistinguishable and the high- and low-resistance values at 0.1 V and 18 V, respectively, remain unchanged throughout the switching cycles (inset in Fig 2C), indicating that the observed switching cannot be attributed to electrical breakdown produced by ionic motion.

Because the resistance of the irradiated LSMO strip is much greater than that of the unirradiated areas (Fig. 1C), most of the voltage drop occurs in this section, thus the switching originates from the irradiated region. It is important to note that even after the irradiated device is switched, the total device resistance is still higher than the resistance of as-grown LSMO, indicating that the irradiated region always dominates the device resistance.

The volatile resistive switching in irradiated LSMO devices (Fig. 2B) resembles the volatile switching commonly observed in IMT materials such as $VO_2$, $V_2O_3$, $V_3O_5$, $SmNiO_3$, $NdNiO_3$, etc.[12,29-32], even though the irradiated LSMO device does not exhibit an insulator-to-metal transition in the low-voltage resistance-temperature measurements (Fig. 1C). Fig 3A shows the resistance of the irradiated device as a function of temperature at several applied voltages from 0.1 V to 16 V, extracted from the I-V curves in Fig. 2B. In the 0.1-12 V range, the resistance-temperature curves show insulating behavior, i.e. the resistance decreases with increasing temperature. Above 13 V however, a voltage-induced transition into a metallic phase is found in which the resistance-temperature slope becomes positive. Fig 3B compares the switching power as a function of temperature in devices before and after irradiation. In the as-grown LSMO device, the switching power *increases* with decreasing temperature, consistent with the electrothermally driven switching (i.e. the lower the base temperature, the more power is needed to induce the switching).[10] On the contrary, the switching power in the irradiated device *decreases* with decreasing temperature, suggesting that Joule heating likely plays a minor role in the resistive switching of ion-irradiated LSMO. Overall, the results in Fig. 3 imply that the high-to-low resistance switching in LSMO irradiated by a focused $Ga^+$ ion beam is voltage-driven.

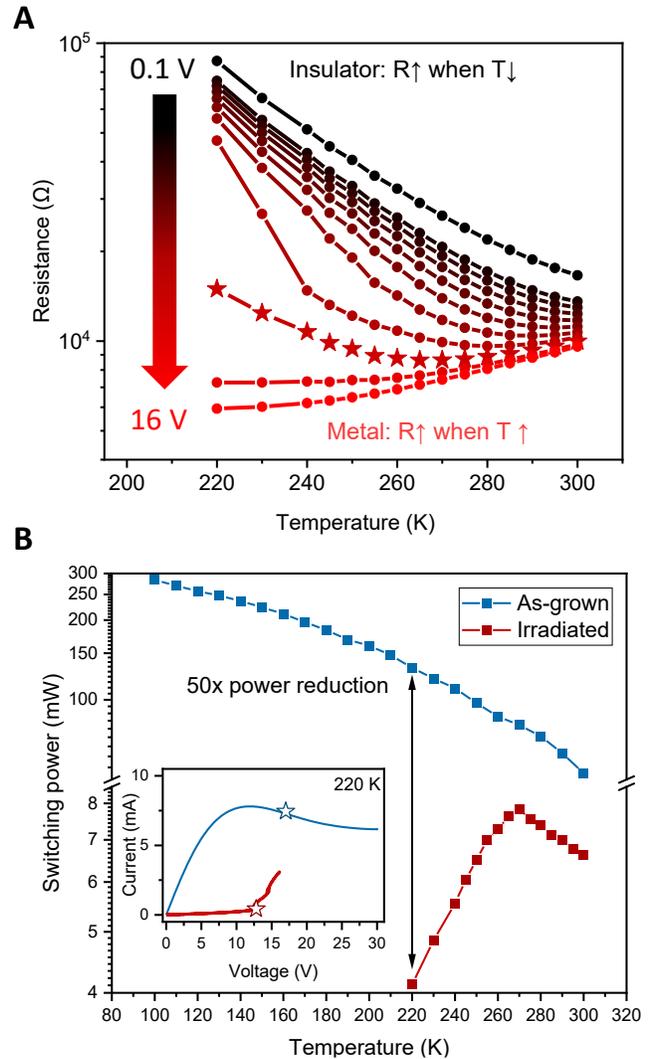

**Figure 3: Nonthermal, voltage-induced insulator to metal switching (A)** Resistance vs. temperature of irradiated LSMO at several applied voltages. The resistance values were extracted from the I-V curves in Fig 2B. Below 12 V, the irradiated device shows insulating behavior. Above 12 V (threshold indicated by stars), metallic behavior is observed. **(B)** Switching power in a pristine device (blue line and symbols) and irradiated device (red and symbols) as a function of temperature. The inset shows current vs. voltage measurement performed at 220 K in the as-grown and irradiated devices. The stars on the V-I curves indicate the moment of switching.

Generally in IMT materials, voltage induced switching suppresses magnetic ordering.[10-12] Since the electrical measurements provided evidence of a voltage induced IMT, it is natural to investigate the effect on the magnetic properties. Surprisingly, we found that voltage-induced switching in irradiated LSMO devices drives the onset of ferromagnetism, as clearly revealed in the anisotropic magnetoresistance (AMR). Figure 4A-C compares the resistance vs. magnetic field when 1 V (below switching) and 15 V (above switching) is applied to the irradiated device. At low voltages (1V), the device shows no significant magnetoresistance. Above the switching threshold (15V), a clear "butterfly-shaped" resistance-field hysteresis loop appears, indicative of ferromagnetism (additional measurements are shown in Suppl. Fig. S1). The observation of ferromagnetic-like magnetoresistance at high voltages in the irradiated devices is in stark contrast with the behavior in the unirradiated LSMO device, where applying high voltages causes significant Joule heating and suppresses ferromagnetism (Suppl. Fig. S2).

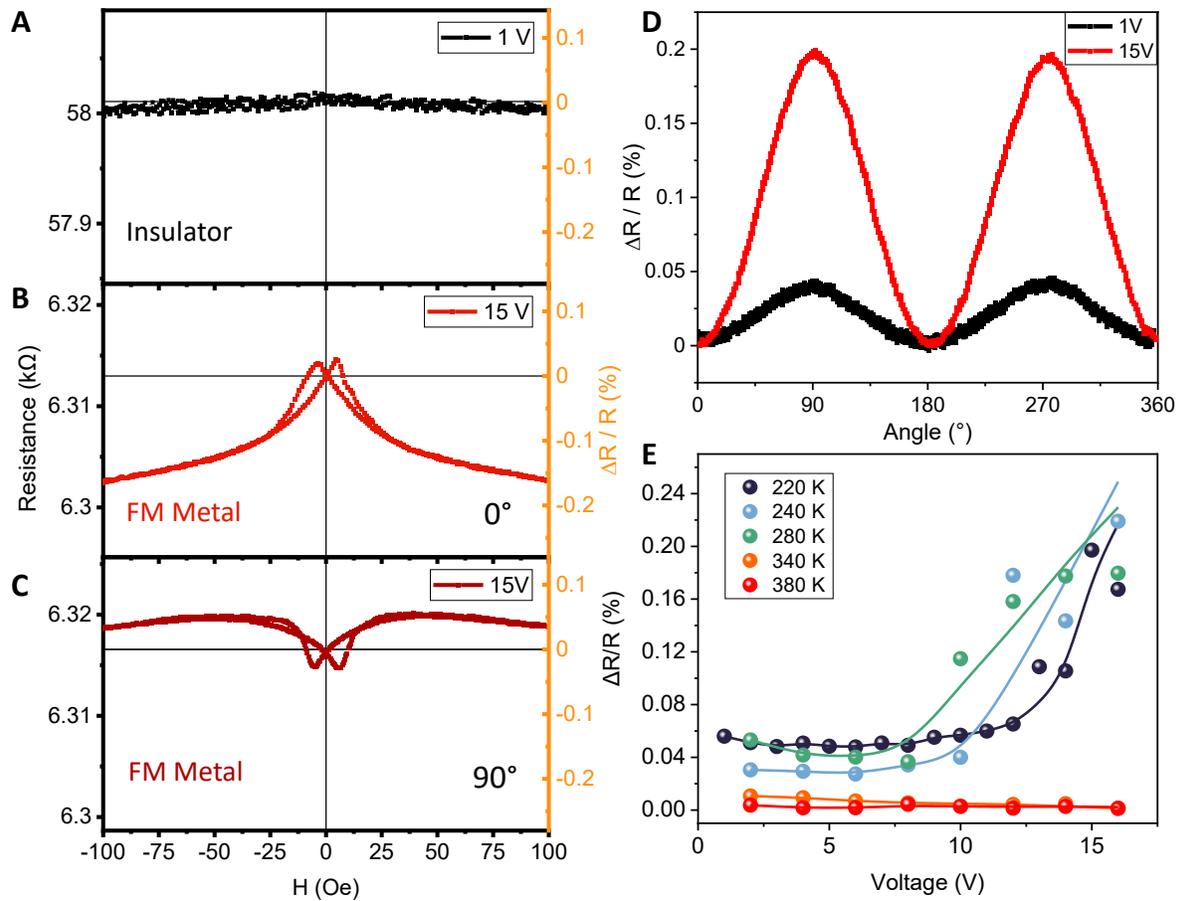

**Figure 4: Voltage-induced onset of magnetic ordering (A)** Top panel shows that magentoresitance is nearly abscent in the irradiated LSMO device at low applied voltages (e.g., at 1 V). Middle and bottom panels **(B,C)** show clear orientation-dependent butterfly hysteretic magentoresistance loops, indicative of AMR. In panel **B**, the magentic field is along the current flow (0º), and in panel **C**, the magnetic field is perpendicular to the current (90º). Measurements were performed at 220 K. **(D)** AMR below (1 V) and above (15 V) the IMT switching voltage treshold. Measurements performed at 220 K in 1 kOe magnetic field. **(E)** AMR magnitude as a function of voltage at several temperatures. At the onset of switching, a sudden jump in the AMR magnitude appears. The AMR magnitude increase only occurs below ~300 K. The lines serve as guides for the eye.

We emphasize again that although the irradiated region is enclosed by pristine LSMO, the resistance of the irradiated region dominates the total device resistance, even after the high-to-low resistance switching. Furthermore, at 220 K, the maximum absolute magnetoresistance of the as-grown device is only ~1 Ω (Suppl. Fig. S2), whereas the irradiated device exhibits an absolute magnetoresistance of ~10 Ω. Thus, our observation of the onset of ferromagnetism above the switching voltage threshold can be directly attributed to the properties of the ion-damaged region.

As resistance-field measurements typically reveal only a fraction of AMR, we performed angle-dependent AMR measurements, revealing the direction of the saturation magnetization in the plane of the film.[33] Fig. 4D compares the AMR in the irradiated LSMO device at 1 V (below switching) and at 15 V (above switching). The AMR curves were recorded using an in-plane 1 kOe field sufficient to saturate the magnetization in any in-plane direction. At 1 V, a small (~0.03%) AMR is found, which increases by a factor of ~6 reaching ~0.2% above the switching threshold. Fig. 4E shows the AMR magnitude as a function of voltage at several temperatures. A threshold-like behavior where the sudden increase of AMR coincides with the high-to-low resistance switching around 12 V. In addition, the switching induced AMR only occurs at temperatures below ~300 K. The temperature dependence of the high voltage AMR indicates that applying voltage to irradiated LSMO provides a qualitatively different response from increasing temperature. This provides interesting, additional evidence that the switching into the ferromagnetic low-resistance state is non-thermal.

**Discussion**

Our experiments directly show that $Ga^+$ ion beam irradiation locally converts metallic LSMO into an insulating phase and applying voltage to the irradiated material induces a novel type of electrical switching into the ferromagnetic metal phase. However, establishing the physical mechanisms responsible for the observed effects remains an open challenge. It has been previously demonstrated that ion irradiation can effectively impact MIT switching behavior in other materials.[34,35] For example, FIB irradiation of $VO_2$, a nonmagnetic MIT oxide, modifies the switching mechanism from one dominated by Joule heating to one primarily driven by non-thermal field-assisted carrier generation.[30,36] The rich phase diagram of LSMO includes antiferromagnetic insulating, ferromagnetic insulating, and ferromagnetic metallic ground states, depending on the hole concertation controlled by the La/Sr ratio[37,38], thus making it an interesting candidate for ion-induced phase manipulation. Previous experiments have shown that ion irradiation in LSMO and related materials can break the Mn-O bonds, create oxygen vacancies and interstitial sites, induce lattice strain, etc., which reduces the $Mn^{3+}/Mn^{4+}$ ratio and stabilizes the low-hole-doping insulating phases[19,39-41]. It is possible that the $Ga^+$ irradiation in our experiments effectively changes the hole concentration, inducing an insulating phase. Because we observed a small AMR in the irradiated samples at voltages below the switching threshold (Fig 4B), this insulating phase could be a charge-ordered ferromagnetic insulator with coexistence of an antiferromagnetic insulator (i.e. phases commonly observed in LSMO with the Sr fraction below 0.2).[42-44] Applying high voltage to the irradiated LSMO could effectively act as electrostatic doping, which increases the hole concentration and drives the material across the IMT, similarly to the effect of increasing the Sr fraction above 0.2. This scenario is also consistent with our observation of large AMR in the voltage-induced metallic state (Fig. 4), as LSMO with the Sr fraction above 0.2 is a ferromagnetic metal in its ground state. Further experiments to probe carrier concertation, $Mn^{3+}/Mn^{4+}$ ratio, and local structural properties are needed to develop a deeper understanding of the insulting ground state and voltage-induced ferromagnetic metal state in $Ga^+$ ion irradiated LSMO.

Another interesting open question relates to the spatial distribution of the electrical IMT switching in the irradiated LSMO device. Commonly, electrically induced volatile IMT switching in materials such as $VO_2$, $V_2O_3$, $V_3O_5$, $NdNiO_3$, $SmNiO_3$, etc., occurs by the formation of a metallic phase filament percolating through an otherwise insulating phase matrix.[11,45-47] Such filamentary switching often exhibits considerable cycle-to-cycle stochasticity in switching voltage threshold, transition time, and resistance values.[48-52] Taking into account that our experiments showed highly deterministic switching characteristics (Fig. 2C), it is possible that the switching in our irradiated LSMO is non-filamentary, similar to the filament-free voltage-induced non-thermal transition in $CaRu_2O_3$.[53] High-resolution imaging experiments that are sensitive to contrast in the electronic metal/insulator and magnetic phases are needed to investigate the spatial distribution of the electrical switching in the irradiated LSMO devices.

**Conclusions**

We showed that low-dose focused $Ga^+$ ion irradiation converts metallic LSMO into an insulator. Irradiated LSMO devices displayed highly repeatable, volatile *insulator-to-metal* resistive switching, contrary to the *metal-to-insulator* switching in the as-synthesized material. The switched metallic phase in the irradiated LSMO device exhibited clear signatures of AMR, showing voltage-induced ferromagnetism. This electrically induced magnetism is in stark contrast to other IMT or MIT materials, in which the electrical phase transition triggering suppresses magnetic ordering. Our experiments further showed evidence that the insulator-to-metal switching in the irradiated devices is non-thermal, allowing for a factor as large as ~50 reduction of the switching power compared to switching in the as-grown material. FIB irradiation thus provides practical means of engineering functional devices in which electronic and magnetic states can be controlled by voltage biasing. Such combined electronic and magnetic switching functionalities could find applications in novel computing technologies that can process and store information in a scalable, energy efficient fashion using both charge and spin degrees of freedom.

**Methods**

<u>Sample preparation</u>. 20-nm-thick $La_{0.7}Sr_{0.3}MnO_3$ films were epitaxially grown on (001)-oriented $SrTiO_3$ substrates with pulsed laser deposition using a laser fluence of 0.7 J/cm$^2$ and a frequency of 1 Hz. The substrates were held at 700 °C during the growth in 0.3 Torr of oxygen.

After deposition, the films were gradually cooled to room temperature in 300 Torr $O_2$ to maintain proper oxygen stoichiometry. Electrodes of (100 nm Au)/ (20 nm Pd) were fabricated using standard photolithography techniques and e-beam evaporation. The bottom Pd layer was used to minimize contact resistance with the LSMO film. 50×100 $\mu m^2$ devices were patterned between electrodes using Ar ion milling. After pristine samples had been characterized, a 1-μm-wide strip stretching across the entire device width was irradiated using a 30 keV $Ga^+$ focused ion beam in a commercial scanning electron microscope.

Electrical measurements. Electrical transport and resistive switching measurements were performed in a Montana Instruments Cryostation s50 using a Keithley 2450 source meter operating in voltage-controlled mode.

Magnetotransport measurements. Magnetotransport measurements were performed in a Quantum Design PPMS Dynacool cryostat using a Keithley 2450 source meter. Anisotropic magnetoresistance measurements were performed using a horizontal rotator which rotates the sample in-plane relative to a 1 kOe magnetic field.

13  Hormoz, S. & Ramanathan, S. Limits on vanadium oxide Mott metal–insulator transition field-effect transistors. *Solid-State Electronics* **54**, 654-659, doi:10.1016/j.sse.2010.01.006 (2010).
14  Torres, F., Basaran, A. C. & Schuller, I. K. Thermal management in neuromorphic materials, devices, and networks. *Advanced Materials* **35**, 2205098, doi: 10.1002/adma.202205098 (2023).
15  Zhou, Y. & Ramanathan, S. Mott memory and neuromorphic devices. *Proceedings of the IEEE* **103**, 1289-1310, doi:10.1109/JPROC.2015.2431914 (2015).
16  Hoffmann, A. *et al.* Quantum materials for energy-efficient neuromorphic computing: Opportunities and challenges. *APL Materials* **10**, doi:10.1063/5.0094205 (2022).
17  Ziegler, J. F., Ziegler, M. D. & Biersack, J. P. SRIM–The stopping and range of ions in matter (2010). *Nuclear Instruments and Methods in Physics Research Section B: Beam Interactions with Materials and Atoms* **268**, 1818-1823, doi: 10.1016/j.nimb.2010.02.091 (2010).
18  Urushibara, A. *et al.* Insulator-metal transition and giant magnetoresistance in $La_{1-x}Sr_xMnO_3$. *Physical Review B* **51**, 14103-14109, doi:10.1103/PhysRevB.51.14103 (1995).
19  Cao, L. *et al.* Metal–insulator transition via ion irradiation in epitaxial $La_{0.7}Sr_{0.3}MnO_{3-\delta}$ thin films. *physica status solidi (RRL)–Rapid Research Letters* **15**, 2100278, doi:10.1002/pssr.202100278 (2021).
20  Pallecchi, I. *et al.* Investigation of FIB irradiation damage in $La_{0.7}Sr_{0.3}MnO_3$ thin films. *Journal of Magnetism and Magnetic Materials* **320**, 1945-1951, doi:10.1016/j.jmmm.2008.02.171 (2008).
21  Takamura, Y., Chopdekar, R., Arenholz, E. & Suzuki, Y. Control of the magnetic and magnetotransport properties of $La_{0.67}Sr_{0.33}MnO_3$ thin films through epitaxial strain. *Applied Physics Letters* **92**, 162504, doi:10.1063/1.2908051 (2008).
22  Wang, B. *et al.* Oxygen-driven anisotropic transport in ultra-thin manganite films. *Nature Communications* **4**, 2778 doi:10.1038/ncomms3778 (2013).
23  Huijben, M. *et al.* Critical thickness and orbital ordering in ultrathin $La_{0.7}Sr_{0.3}MnO_3$ films. *Physical Review B* **78**, 094413, doi:10.1103/PhysRevB.78.094413 (2008).
24  López, A. *et al.* Electrolyte Gated Synaptic Transistor based on an Ultra-Thin Film of La0.7Sr0.3MnO3. *Advanced Electronic Materials* **9**, 2300007, doi: 10.1002/aelm.202300007 (2023).
25  Manca, N., Pellegrino, L. & Marré, D. Reversible oxygen vacancies doping in $La_{0.7}Sr_{0.3}MnO_3$ microbridges by combined self-heating and electromigration. *Applied Physics Letters* **106**, 203502, doi:10.1063/1.4921342 (2015).
26  Del Valle, J., Ramírez, J. G., Rozenberg, M. J. & Schuller, I. K. Challenges in materials and devices for resistive-switching-based neuromorphic computing. *Journal of Applied Physics* **124**, 211101, doi:10.1063/1.5047800 (2018).
27  Chen, Y., Ziese, M. & Esquinazi, P. Bistable resistance state induced by Joule self-heating in manganites: A general phenomenon. *Applied physics letters* **88**, 222513 doi:10.1063/1.2209204 (2006).
28  Balcells, L. *et al.* Electroresistance and Joule heating effects in manganite thin films. *Journal of Applied Physics* **113**, 073703 doi:10.1063/1.4792222 (2013).
29  del Valle, J. *et al.* Subthreshold firing in Mott nanodevices. *Nature* **569**, 388-392, doi:10.1038/s41586-019-1159-6 (2019).
10

**Acknowledgements**

This work was supported as part of the Quantum Materials for Energy Efficient Neuromorphic Computing (Q-MEEN-C), an Energy Frontier Research Center funded by the U.S. Department of Energy, Office of Science, Basic Energy Sciences under Award # DE-SC0019273. This work was performed in part at the San Diego Nanotechnology Infrastructure (SDNI) of UCSD, a member of the National Nanotechnology Coordinated Infrastructure, which is supported by the National Science Foundation (Grant ECCS-2025752).

**Author Contributions**

N.G. and D.S. fabricated the devices. N.G. performed the measurements and analyzed the data with input from P.S and I.K.S. I.K.S. and Y.T. supervised the design, execution and data analysis of the experiment. All authors contributed to the interpretation of the results and production of the paper.

**Competing interests**

The authors declare no competing interests.

# Supplementary Materials for

**Electrically induced ferromagnetism in an irradiated complex oxide**

Nareg Ghazikhanian, Pavel Salev, Dayne Sasaki, Yayoi Takamura, Ivan K. Schuller

Corresponding author: nghazikh@ucsd.edu

**The supplementary information includes:**

Figs. S1 to S2

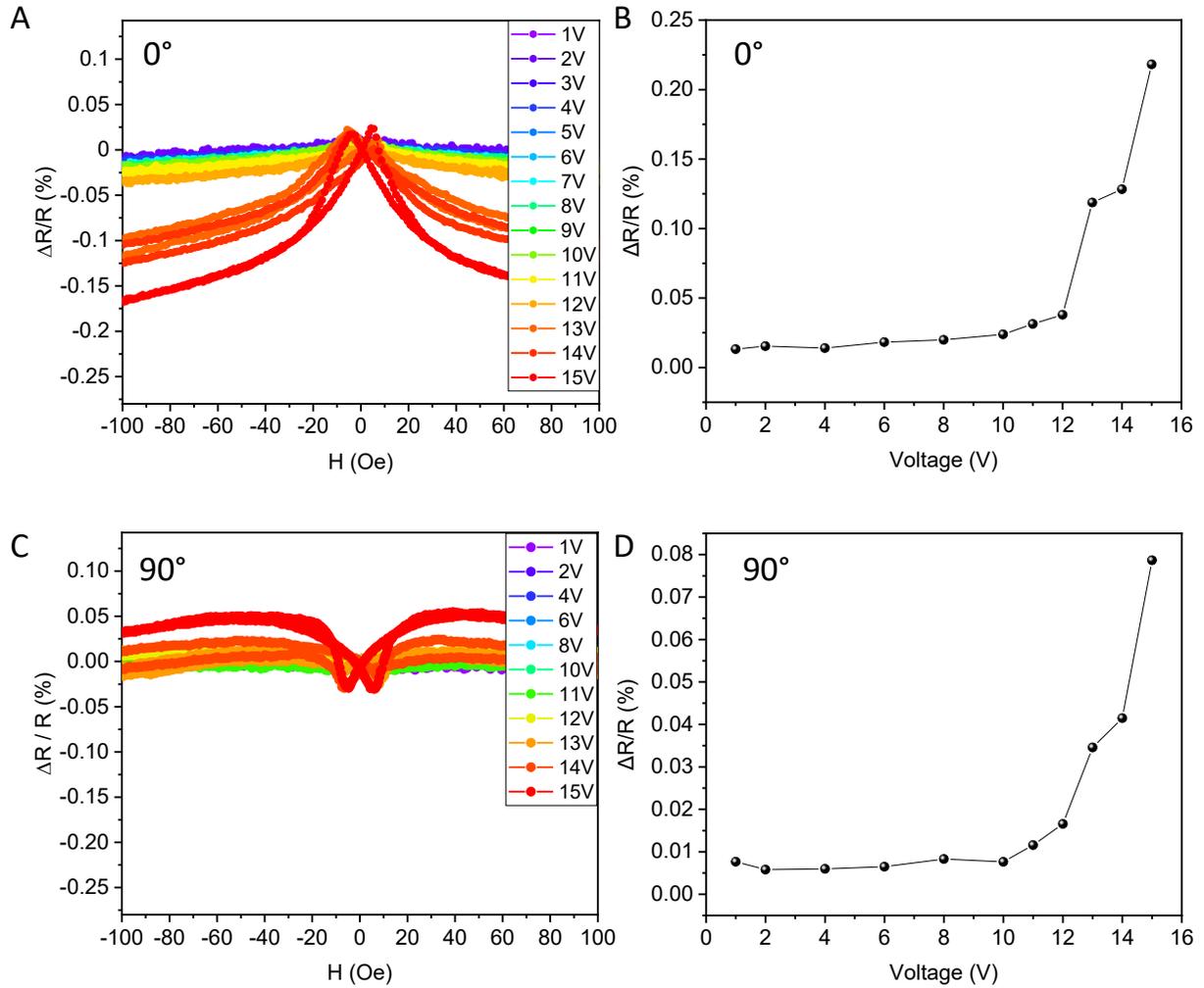

Figure S1. Additional magnetoresistance measurments in irradiated device (A,C) Magnetoresistance measurements as a function of appled voltage performed in the irradiated LSMO device with the current parallel to the magnetic field (A) and current perpendicular to the magnetic field (C). (B,D) Amplitude of change in magnetoresistance for the irradiated LSMO device when current is parrallel to the magnetic field (B) and perpendicular to the magnetic field (D). For both sets of measurements, a sudden increase in magnetoresistance is observed above ~12 V.

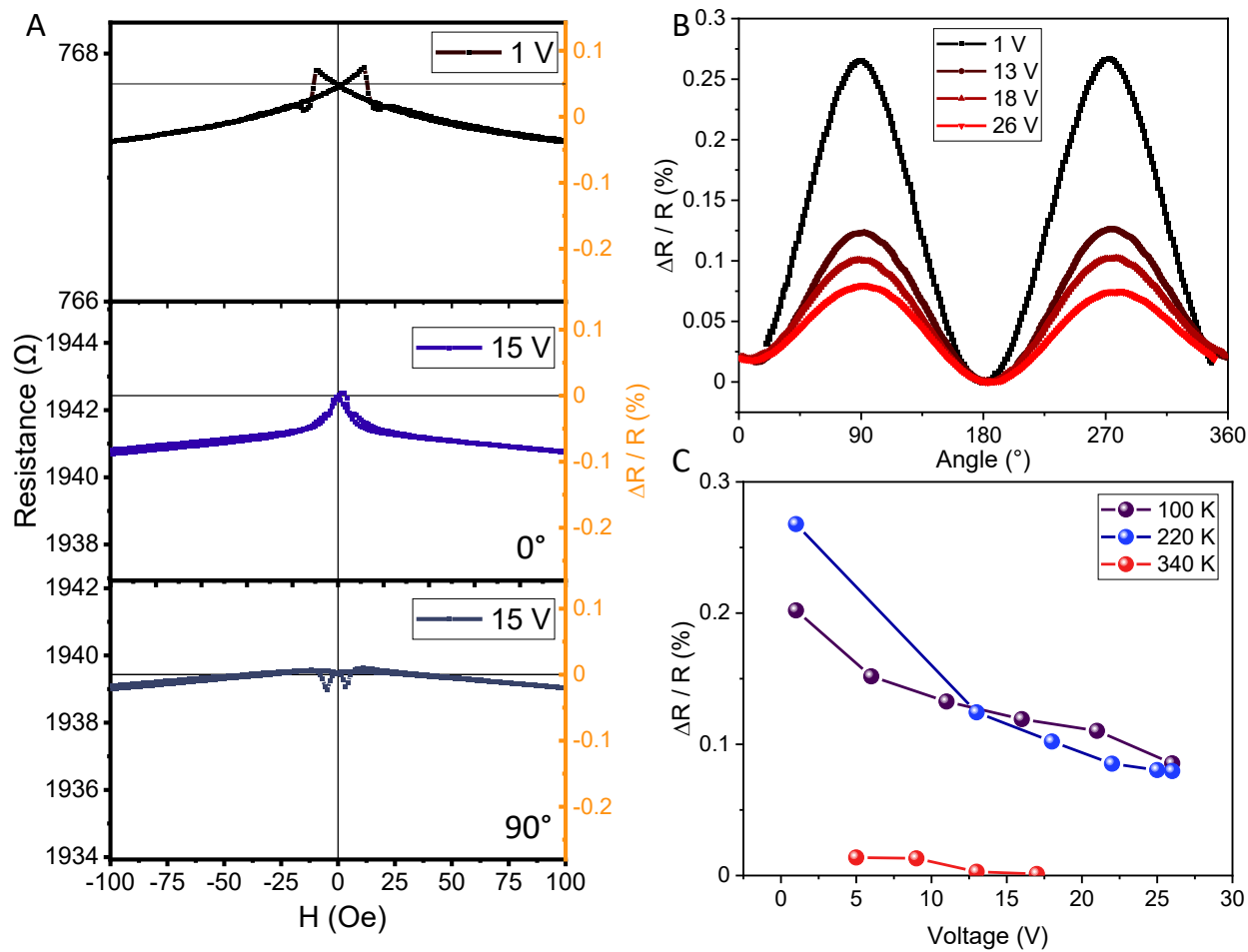

**Figure S2. Magnetoresistance measurements for as-grown LSMO (A)** Magnetoresistance measurements at 220K for the as-grown device at different values of applied voltage and device orientation. At low values of applied voltage (V= 1 V), the device shows prominent butterfly hysteretic peaks commonly associated with ferromagnetism. At larger values of applied voltage (V=15 V), these peaks are diminished, indicating a decrease in the ferromagnetic order. The maximum magnitude of resistance change was observed to be only ~1 Ω. **(B)** AMR at 220K in 1 kOe for different applied voltages. **(C)** Amplitude of AMR as a function of temperature. Over the entire temperature range, increasing applied voltage surpresses the amplitude of AMR.